\documentclass[10pt]{article}
\usepackage{multicol}
\usepackage{graphicx}
\usepackage{amsmath}
\usepackage[a4paper]{geometry}
\usepackage{hyperref}
\usepackage{rotating}

\begin{document}

\begin{center}
{\Large \bf Study of dependence of kinetic freezeout temperature
on production cross-section of the particles in various centrality
intervals in Au-Au and Pb-Pb collisions at high energies}

\vskip1.0cm

M.~Waqas$^{1,}${\footnote{Corresponding author. Email (M.Waqas):
waqas\_phy313@yahoo.com; waqas\_phy313@ucas.ac.cn}},G. X.
Peng$^{1,2}$ {\footnote{Corresponding author. Email (G. X. Peng):
gxpeng@ucas.ac.cn}}
\\

{\small\it  $^1$ School of Nuclear Science and Technology,
University of Chinese Academy of Sciences, Beijing 100049, China,

$^2$ Theoretical Physics Center for Science Facilities, Institute
of High Energy Physics, Beijing 100049, China}

\end{center}

{\bf Abstract:} Transverse momentum spectra of $\pi^+$, $p$,
$\Lambda$, $\Xi$ or $\bar\Xi^+$, $\Omega$ or $\bar\Omega^+$ and
deuteron ($d$) in different centrality intervals in
nucleus-nucleus collision at center of mass energy are analyzed by
the blast wave model with Boltzmann Gibbs statistics. We extracted
Kinetic freeze out temperature, transverse flow velocity and
kinetic freezeout volume from the transverse momentum spectra of
the particles. It is observed that the non-strange and strange
(multi-strange) particles freezeout separately due to different
reaction cross-sections. While the freezeout volume and transverse
flow velocity are mass dependent.  They decrease with the rest
mass of the particles. The present work reveals the scenario of
double kinetic freezeout in nucleus-nucleus collisions.
Furthermore, the kinetic freezeout temperature and freezeout
volume are larger in central collisions than the peripheral
collisions. However the transverse flow velocity remains almost
unchanged from central to peripheral collisions.
\\



\vskip1.0cm

\begin{multicols}{2}

{\section{Introduction}}
Freezeout stages are very important
because they provide essential information about the emission of
the particles at those stages. Generally, there are two freezeout
stages found in literature namely chemical freezeout and kinetic
freezeout stage and both of them correspond to their respective
temperatures. The chemical freezeout is the intermediate stage in
high energy collisions where the intra-nuclear collisions among
the particles are inelastic as well as the ratio of various types
of particles remain unchanged, and the temperature of the
particles at this stage is chemical freezeout temperature that
describes the excitation degree of the system at the chemical
freezeout stage. Correspondingly, the thermal/kinetic freezeout is
the last but not the least stage in high energy collisions. At
this stage, the intra-nuclear collisions among the particles are
elastic. The transverse momentum distributions of various kinds of
particles are no longer changed at thermal freezeout stage, and
the temperature at this stage is called kinetic freezeout
temperature.

According to thermal and statistical model [1--4], the chemical
freezeout temperature ($T_{ch}$) in central nucleus- nucleus
collisions increases with increase of the collisions energy from a
few GeV to above 10 GeV and then saturates in an energy range from
more than a dozen of GeV. At top Relativistic Heavy Ion Collider
(RHIC) and Large Hadron Collider (LHC), the maximum $T_{ch}$ is
160 MeV, although there is a slight decrease from RHIC to LHC
energy, but the situation of kinetic freezeout temperature ($T_0$)
is complex. At first, $T_0$ in central collisions increases with
increasing the collision energy from a few GeV to above 10 GeV,
but this tendency can either be saturated, decrescent or
increscent. On the other hand, $T_{ch}$ in central nucleus-nucleus
collisions is a little larger than in peripheral nucleus-nucleus
collisions, however, there are three possible trends of $T_0$ from
central to peripheral collisions which are; (1) $T_0$ increase
from central to peripheral collisions, (2) $T_0$ decrease from
central to peripheral collisions, and (3) $T_0$ remains constant
from central to peripheral collisions. It is very important to
search for the correct trend of $T_0$ with energy and centrality.
Furthermore, there are different kinetic freezeout scenarios found
in literature which include single, double, triple and multiple
kinetic freezeout scenarios [5--10]. In single kinetic freezeout
scenario, one set of parameters is used for the strange,
multi-strange and non-strange particles. In double kinetic
freezeout scenario, one set of parameters is used for strange
(multi-strange) and other for non-strange particles, separate sets
of parameters are used for strange, multi-strange and non-strange
particles in triple kinetic freezeout scenario. While in multiple
kinetic freezeout scenario, separate sets of parameters are used
for each particle. The trend of transverse flow velocity
($\beta_T$) and freezeout volume ($V$) with energy is an
increasing trend in most of the literatures [6, 11-16]. Most of
the literature claims the decreasing (or invariant) trend of
$\beta_T$ and $V$ from central to peripheral collisions [10, 15,
16--18].

 The transverse momentum spectra ($p_T$) of the particles are very important observables due to the reason that
 they provide very essential information about the equilibrium dynamics and isotropy of the system in high energy
 collisions [9]. In present work, we will analyze the $p_T$ spectra of $\pi^+$, $p$, $\Lambda$, $\Xi$ ($\bar \Xi^+$),
 $\Omega$ ($\bar \Omega^+$) and deuteron ($d$) in nucleus-nucleus collisions at center of mass energy.

The remainder of the paper consists of the method and formalism,
and results and discussion in section 2 and 3 respectively, and
the summary of our main observations and conclusions are presented
in section 4.
\\

{\section{The method and formalism}} There are various models
suggested for the extraction of $T_0$, $V$ and $\beta_T$, e:g
blast wave model with Boltzmann Gibbs statistics (BGBW) [19--21],
blast wave model with Tsallis statistics [TBW] [22--24], an
alternative method by using Tsallis statistics [25--31] and an
alternative method by using blast wave model with Boltzmann Gibbs
statistics [32--37]. In this work, we choose the blast wave model
with Boltzmann Gibbs statistics which is phenomenological model
and is used for the spectra of hadrons based on the flow of local
thermal sources with global variables of temperature, volume and
transverse flow velocity.

According to reference [38--40], the $p_T$ distribution of the
BGBW can be written as
\begin{align}
f(p_T)=&\frac{1}{N}\frac{dN}{dp_T} =C \frac{gV}{(2\pi)^2} p_T m_T \int_0^R rdr \nonumber\\
& \times I_0 \bigg[\frac{p_T \sinh(\rho_1)}{T_0} \bigg] K_1
\bigg[\frac{m_T \cosh(\rho_1)}{T_0} \bigg],
\end{align}
where $C$ stands for the normalization constant, $g$ represents
the degeneracy factor of the particles, $V$ is the freezeout
volume, $m_T=\sqrt{p_T^2+m_0^2}$ is the transverse mass ($m_0$ is
the rest mass of the particle), r is the radial coordinate, $R$ is
the maximum $r$, $\rho=\tanh^{-1} [\beta(r)]$ is the boost angle,
$\beta(r)=\beta_S(r/R)^{n_0}$ is a self-similar flow profile,
$\beta_S$ is the flow velocity on the surface, as mean of
$\beta(r)$, $\beta_T=(2/R^2)\int_0^R r\beta(r)dr=2\beta_S/(n_0+2)$
if $n_0$=2, $\beta_T$=0.5$\beta_S$, because the maximum $\beta_S$
is 1c and the maximum value of $\beta_T$ is 0.5, but if $n_0$=1,
then it will result in $\beta_T=(2/3)\beta_S$, hence the maximum
$\beta_T$ is (2/3)c. However if one use $n_0$ as a free parameter
[41], it fluctuates largely by several times increases 1 in the
number of free parameters. $I_0$ and $K_1$ are the Bessel modified
functions of the first and second kind respectively.

Eqn. (1) is not enough for the description of the whole $p_T$
spectra, particularly, when the maximum $p_T$ reaches to 100 GeV/c
at LHC collisions [42], where several $p_T$ regions [43] has been
observed by the model analysis. These regions include the first
$p_T$ region with $p_T$ $<$4.5 GeV/c, the second and third region
with 4-6 GeV/c $<$ $p_T$ $<$ 17-20 GeV/c and $p_T$ $>$ 17-20 GeV/c
respectively. It is expected that different $p_T$ regions are
corresponding to different interacting mechanisms, such as the
effects and changes by medium, nuclear transparency and the effect
of number of strings etc, which are discussed in detail in [17].
Therefore, for the complete description of the whole $p_T$, we can
use the functions such as Tsallis Levy [44, 45], the Hagedorn
function [42, 46, 47] to the spectra in high and very high $p_T$
regions and it corresponds to an the inverse power law. In this
work, we used the inverse power law to describe the $p_T$ spectra
in high $p_T$ regions, that is
\begin{align}
f_H(p_T)=&\frac{1}{N}\frac{\mathrm{d}N}{\mathrm{d}p_\mathrm{T}}=
Ap_T \bigg( 1+\frac{p_T}{p_0} \bigg)^{-n},
\end{align}
where $N$ and $A$ represents the number of particles and
normalization constant respectively, and $p_0$ and $n$ are the
free parameters. There are several modified versions of Hagedorn
function found in literature [48--54].

Generally, the two main process responsible for the contribution
of $p_T$ spectra are soft excitation (contributes soft component
in low $p_T$ region) and hard scattering process (contributes in
whole $p_T$ region). Eqn. (1) is taken into account for the soft
excitation process and eqn.(2) for the hard scattering process.
Eqn. (1) and (2) can be superposed by two methods i.e; (1) super
position principle, where the contribution regions of components
overlap each other and (2) by Hagedorn model (usual step
function), when there is no overlapping of different regions of
different components. According to the first method

\begin{align}
f_0(p_T)=kf_S(p_T)+(1-k)f_H(p_T),
\end{align}
where $k$ represents the contribution fraction of the first
component and (1-k) represents the contribution function of the
second component.

The usual step function can be used to structure the superposition
of the Eqn. (1) and (2). According to Hagedorn model [42, 46, 47],
the usual step function can also be used for the superposition of
the two functions, as
\begin{align}
f_0(p_T)=A_1\theta(p_1-p_T) f_1(p_T) + A_2
\theta(p_T-p_1)f_2(p_T),
\end{align}
where $A_1$ and $A_2$ are the fraction constants  which give the
two components to be equal to each other at $p_T$=$p_1$.

 It should be noted that soft and hard components in Eq. (3) and
(4) are treated in different ways in the whole $p_T$ region. Eq.
(3) is used for the contribution of soft component in the range
0-2$\sim$3 GeV/c or a little more. However in case of the
contribution of hard component, even though the main contribution
in low $p_T$ region is soft excitation process, but it covers the
whole $p_T$ region. In Eq. (4), in the range from 0 to $p_1$ and
from $p_1$ up to maximum are the contributions of soft and hard
components respectively, and there is no mixed region for the two
components. In addition, we would like to point out that in the
present work we have used Eq. (1) (which is a single component
BGBW) only, but Eq. (3) and (4) are stated in order to present the
whole methodology and treatment (if Eq. (2) is used). If we use
double component BGBW, then we can use either eqn. (3) or (4) to
combine the two components.
//

{\section{Results and discussion}} Figure 1 demonstrates the
transverse momentum ($p_T$) spectra, [(1/2$\pi$$p_T$)
$d^2$$N$/$dyd$$p_T$] or [$1/N_{ev}$(1/2$\pi$$p_T$)
$d^2$$N$/$dyd$$p_T$] of $\pi^+$, $p$, $\Lambda$, $\bar \Xi^+$,
$\bar \Omega^+$ and deuteron $(d)$ in various centrality classes
in Au-Au collisions at 62.4 GeV. The spectra is distributed in
different centrality classes, e:g for $\pi^+$ and $p$, 0--5\%,
5--10\%, 10--20\%, 20--30\%, 30--40\%, 40--50\%, 50--60\%,
60--70\% and 70--80\%, for $\Lambda$, 0--5\%, 5--10\%, 10--20\%,
20--30\%, 30--40\%, 40--60\% and 60--80\%, for $\bar \Xi^+$,
0--5\%, 5--10\%, 10--20\%, 20--40\%, 40--60\% and 60--80\%, for
$\bar \Omega^+$, 0--20\%, 20--40\% and 40--60\% at $|y|<0.1$, and
for deuteron ($d$), 0--10\%, 10--20\%, 20--40\%, 40--60\% and
60--80\%, at $|y|<0.3$. The symbols are cited from the
experimental data measured by the STAR Collaboration at
Relativistic Heavy Ion collider (RHIC) [21, 55, 56]. In the
figure, the curves are our fitted results with Eq. (1). The
corresponding values of free parameters ($T_0 $, $V$, $\beta_T$
and $n_0$), normalization constant $(N_0)$, $\chi^2$ and number of
degree of freedom (ndof) are listed in Table 1, in which the
parameter trend will be analyzed and discussed later in this
section. One can see that the $p_T$ spectra of the particles are
shown to obey approximately the blast wave model with Boltzmann
Gibbs statistics. Furthermore, the spectra of $\pi^+$ in 5--10\%,
10--20\%, 20--30\%, 30--40\%, 40--50\%, 50--60\%, 60--70\% and
70--80\% centrality intervals are scaled with 1/3, 1/7, 1/18,
1/40, 1/100, 1/250, 1/700 and 1/1500 respectively, while the
centrality intervals 5--10\%, 10--20\%, 20--30\%, 30--40\%,
40--50\%, 50--60\%, 60--70\% and 70--80\% of $p$ are scaled by
1/3, 1/7, 1/26, 1/60, 1/150, 1/250, 1/400 and 1/600 respectively.
\begin{figure*}[htb!]
\begin{center}
\hskip-0.153cm
\includegraphics[width=15cm]{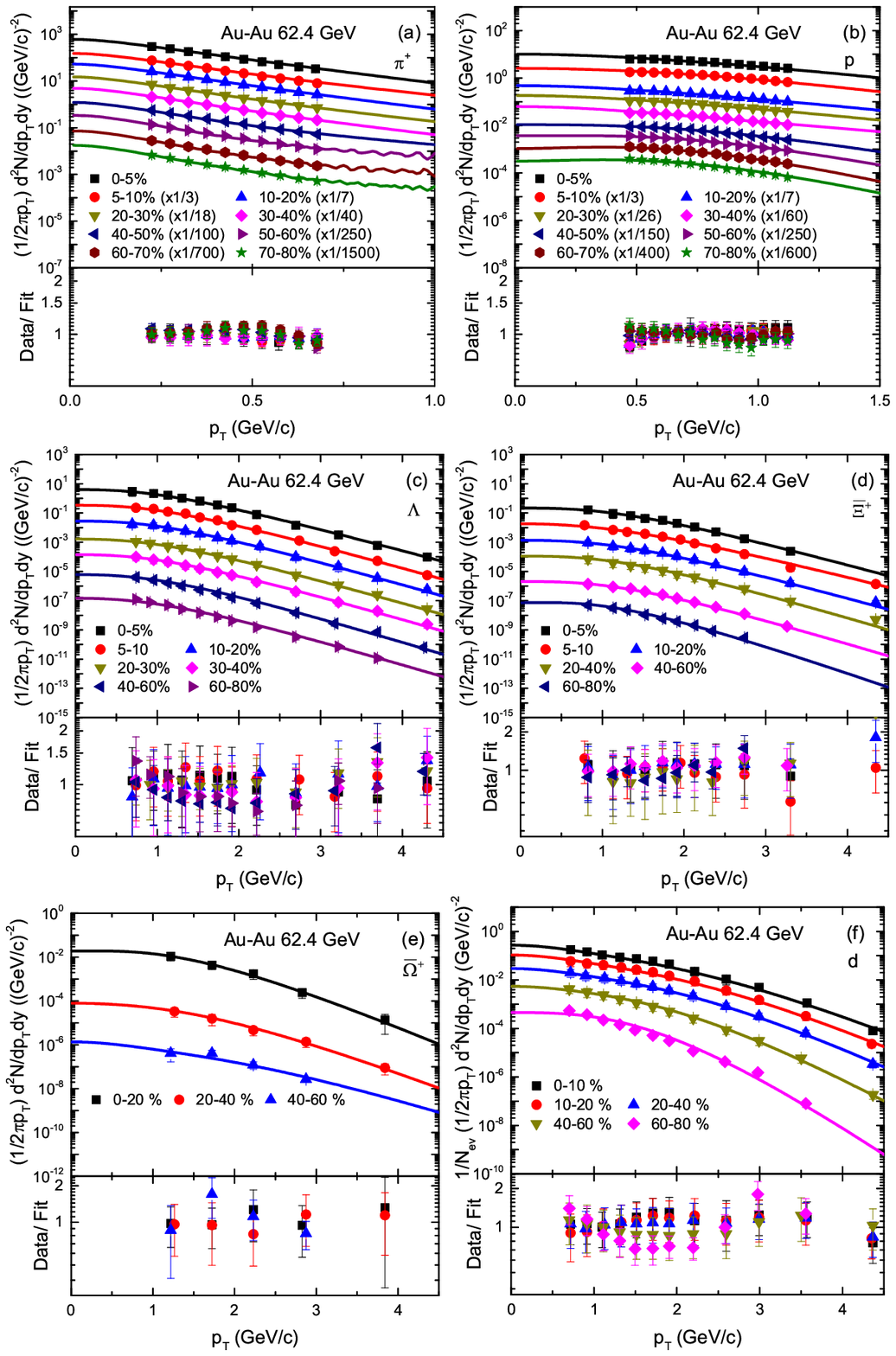}
\end{center}
Fig. 1. Transverse momentum spectra of $\pi^+$, $p$, $\Lambda$,
$\bar \Xi^+$ and $\bar \Omega^+$ rapidity at $|y|<0.1$, and
deuteron $(d)$ at rapidity $|y|<0.3$, produced in different
centrality intervals in Au-Au collisions at 62.4 GeV. Different
symbols represent the $p_T$ spectra of different particles
measured by the STAR collaboration [21, 55, 56] and the curves are
our fitted results with BGBW model. The corresponding results of
data/fit is presented in each panel.
\end{figure*}
\begin{figure*}[htb!]
\begin{center}
\hskip-0.153cm
\includegraphics[width=15cm]{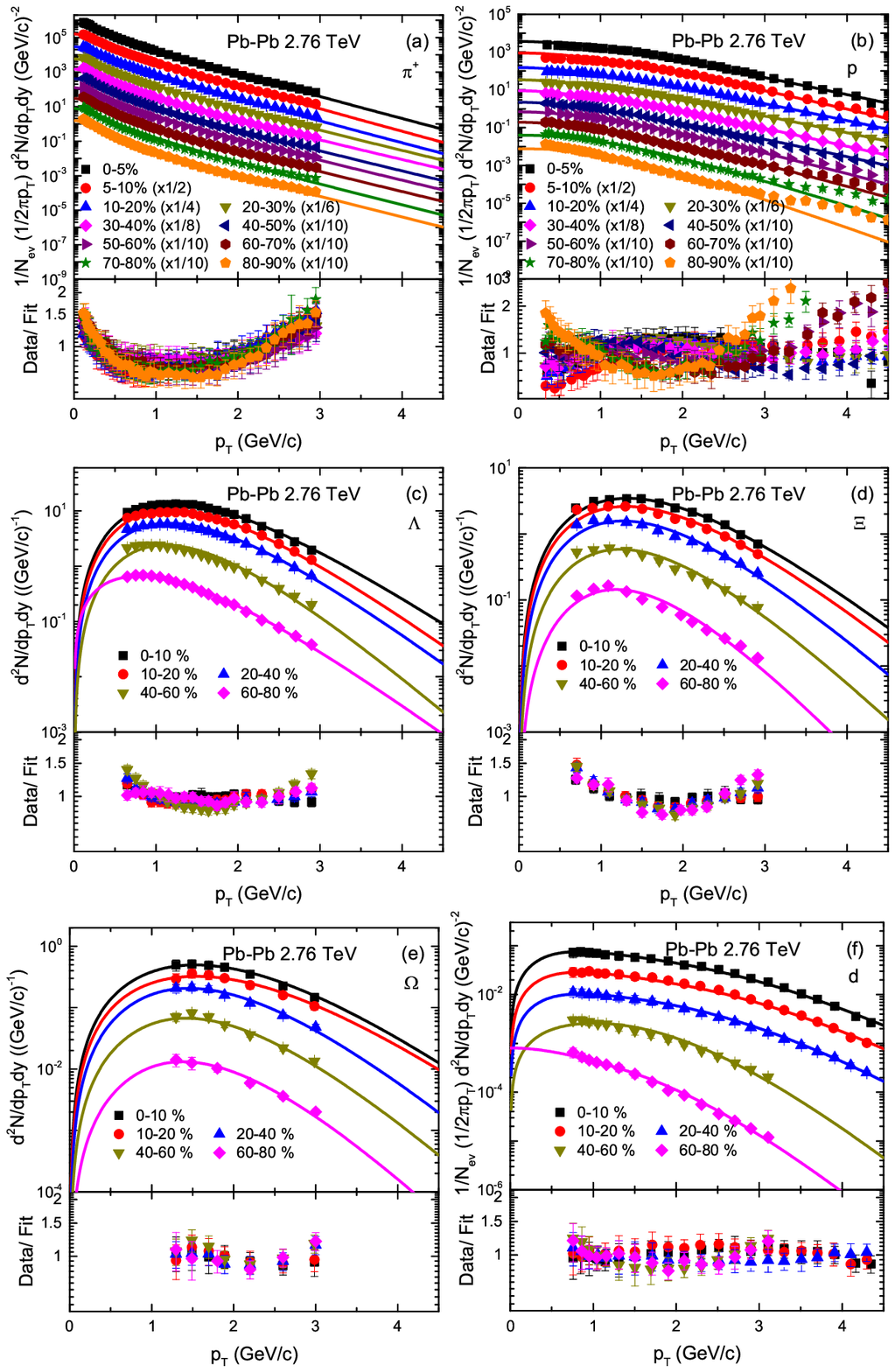}
\end{center}
Fig. 2. Transverse momentum spectra of $\pi^+$, $p$, $\Lambda$,
$\Xi$, $\Omega$ and deuteron $(d)$ produced in different
centrality intervals in Pb-Pb collisions at 2.76 TeV at rapidity
$|y|<0.5$. Different symbols represent the $p_T$ spectra of
different particles measured by the ALICE collaboration [57--59]
and the curves are our fitted results with BGBW model. The
corresponding results of data/fit is presented in each panel.
\end{figure*}

Figure 2 is similar to Fig.1, but it shows the the $p_T$ spectra
of $\pi^+$, $p$, $\Lambda$, $\Xi$, $\Omega$ and deuteron $(d)$ in
different centrality intervals in Pb-Pb collisions at 2.76 TeV.
The spectra is distributed in different centrality intervals, e:g
for $\pi^+$, and $p$; 0--5\%, 5--10\%, 10--20\%, 20--30\%,
30--40\%, 40--50\%, 50--60\%, 60--70\% 70--80\% and 80--90\% at
$|y|<0.5$, for $\Lambda$, $\Xi$, and $\Omega$; 0--10\%, 10--20\%,
20--40\%, 40--60\%, and 60--80\%, for $\Omega$; 0--10\%, 10--20\%,
20--40\%, 40--60\% and 60--80\% at $y=0$, and for deuteron ($d$);
0--10\%, 10--20\%, 20--40\%, 40--60\% and 60--80\%, at $|y|<0.5$.
The spectra of $\pi^+$ and $p$ in 5--10\%, 10--20\%, 20--30\%,
30--40\%, 40--50\%, 50--60\%, 60--70\% and 70--80\% centrality
intervals are scaled with 1/2, 1/4, 1/6, 1/8, 1/10, 1/10, 1/10,
1/10 and 1/10 respectively. The symbols are cited from the
experimental data measured by the ALICE Collaboration at Large
Hadron Collider (LHC) [57--59]. In the figure, the curves are our
fitted results with Eq. (1). The corresponding values of free
parameters ($T_0 $, $V$, $\beta_T$ and $n_0$), normalization
constant $(N_0)$, $\chi^2$ and number of degree of freedom (ndof)
are listed in Table 1, in which the parameter trend will be
analyzed and discussed below. One can see that the $p_T$ spectra
of the particles are shown to obey approximately the blast wave
model with Boltzmann Gibbs statistics. We comment here by the way
that we have used the method of least square to obtain the
parameters in present work, and the fits (especially the ALICE
data) to BGBW model are not good for quite abundant hadron
species, such as $\pi^+$ and protons due to the generation of
non-inclusion resonance in low $p_T$ region. In addition, we would
also like to point out that the values of $\chi^2$ varies
especially in some cases in central collisions increases due to
poor fitting.
\begin{figure*}[htb!]
\begin{center}
\hskip-0.153cm
\includegraphics[width=15cm]{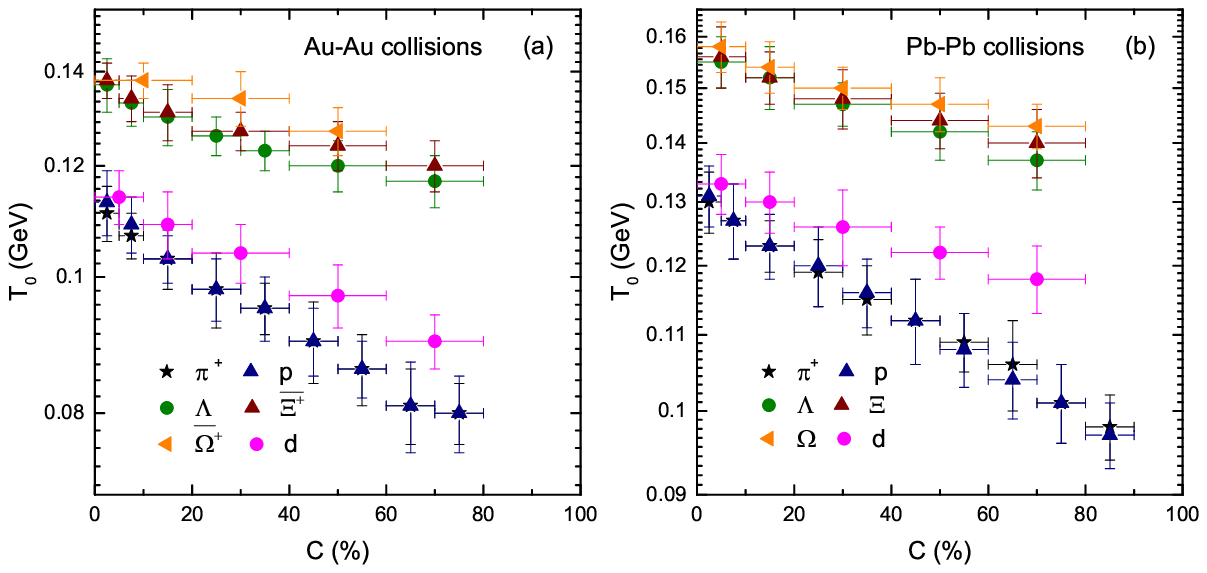}
\end{center}
Fig. 3. Dependence of $T_0$ on centrality class ($C\%$) and rest
mass ($m_0$)  of the particle.
\end{figure*}

Figure 3 analyze the dependence of kinetic freezeout temperature
($T_0$) on centrality class ($C\%$) and mass of the particles.
Panel (a) and (b) show the result for Au-Au and Pb-Pb collisions
respectively. The colored symbols represent different species of
particles and the particles from left to right shows the result of
$T_0$ from central to peripheral collisions. One can see that the
kinetic freezeout temperature of the emission source decreases
with the decrease of centrality from central to peripheral
collisions. The central collision corresponds to very violent
collision due to large number of participant nucleons that makes
the degree of excitation of system high and results in high
temperature, but as the centrality decreases, the collision become
less and less violent due to small number of particles invloved in
the interaction, which results in decreasing the degree of
excitation of they system and correspondingly the temperature
decreases. This result is consistent with [5, 6, 18, 27, 28, 29,
60], but inconsistent with [61--65]. In addition, the dependence
of $T_0$ on $m_0$ is not clear. The pion and proton have almost
same values for $T_0$, and similarly the strange (muti-strange)
particles have almost same values for $T_0$. Deuteron has the
largest mass and it freezeout at the same time with pion and
proton. The reason maybe the production cross-section of the
interacting particle. According to kinematics, the reactions with
lower cross-section is supposed to be switched-off at higher
temperatures/densities or earlier in time than the reactions with
higher cross-sections. $\pi^+$, $p$ and $d$ are non-strange
particles, so they have the same $T_0$, while $\Lambda$,
$\Xi$($\bar \Xi^+$) and $\Omega$($\bar \Omega^+$) are strange
flavored particles, so they have the same $T_0$. The non-strange
particles have larger production cross-section than the strange or
multi-strange particles, therefore the non-strange particles
freezeout later than the strange (multi-strange) particles. This
result is consistent with one of our recent work [10], however
ref.[10] observed separate decoupling of strange and multi-strange
also. It is noteworthy that the observed $T_0$ at RHIC is lower
than that of LHC. In addition, we would also like to point out
that several previous literature studied the blast wave fit with
different methods and got different results from our recent work.
In the present work, the least square method is used, and we
observed the double kinetic freezeout scenario, while the previous
literature observed single or multiple kinetic freezeout scenario.
\begin{figure*}[htb!]
\begin{center}
\hskip-0.153cm
\includegraphics[width=15cm]{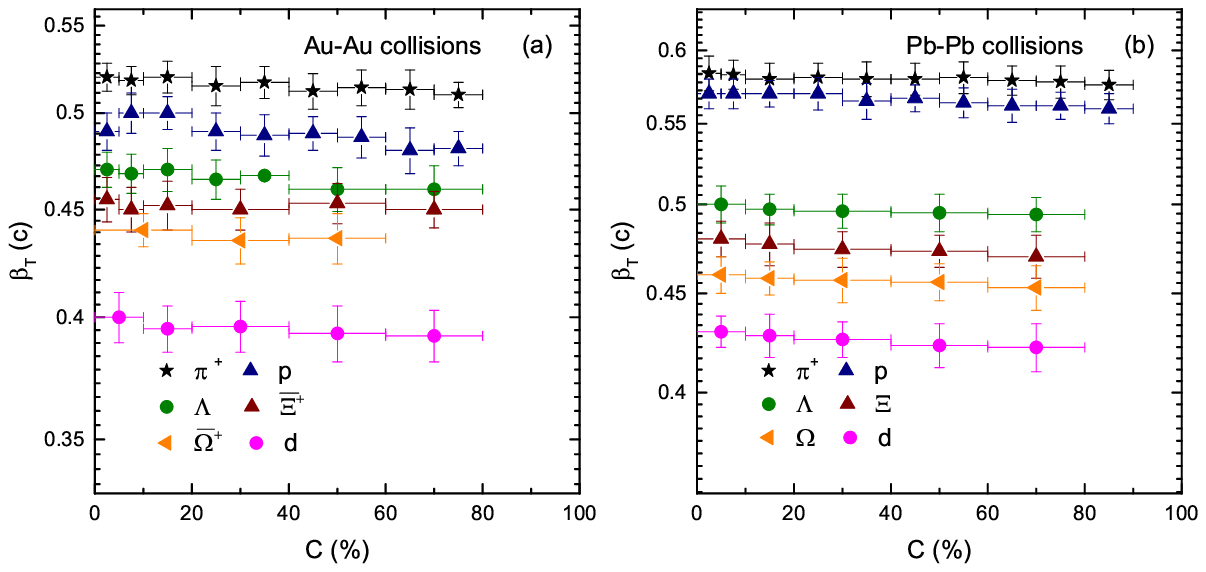}
\end{center}
Fig. 4. Dependence of $\beta_T$ on centrality class ($C\%$) and
rest mass ($m_0$)  of the particle.
\end{figure*}
\begin{figure*}[htb!]
\begin{center}
\hskip-0.153cm
\includegraphics[width=15cm]{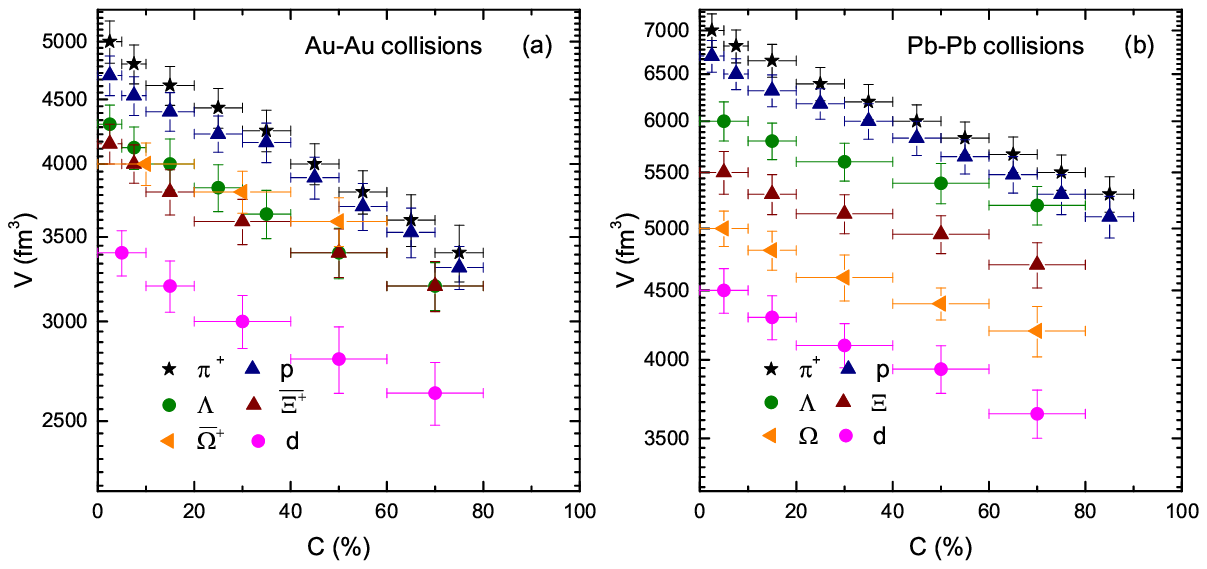}
\end{center}
Fig. 5. Dependence of $V$ on centrality class ($C\%$) and rest
mass ($m_0$)  of the particle.
\end{figure*}

\begin{table*}
{\scriptsize Table 1. Values of free parameters $T_0$ and
$\beta_T$, V and q, normalization constant ($N_0$), $\chi^2$, and
degree of freedom (dof) corresponding to the curves in Figs. 1--2.
\vspace{-.50cm}
\begin{center}
\begin{tabular}{ccccccccccc}\\ \hline\hline
Collisions       & Centrality     & Particle     & $T_0$ (GeV)
& $\beta_T$ (c)    & $V (fm^3)$        & $N_0$              &
$n_0$       &$\chi^2$/dof\\ \hline
   Fig. 1        & $0-5\%$        & $\pi^+$      &$0.111\pm0.005$  & $0.520\pm0.008$  & $5000\pm193$  & $0.25\pm0.06$          & $0.8$             & 3/5\\
   Au-Au         &  $5-10\%$      & --           &$0.107\pm0.004$  & $0.518\pm0.008$  & $4800\pm170$  & $0.24\pm0.004$          &$1.3$             & 7/5\\
   62.4 GeV      & $10-20\%$      & --           &$0.103\pm0.005$  & $0.520\pm0.009$  & $4615\pm165$  & $0.185\pm0.004$         &$2.6$             & 2/5\\
                 & $20-30\%$      & --           &$0.098\pm0.006$  & $0.515\pm0.011$  & $4430\pm161$  & $0.136\pm0.0005$        &$1.3$             & 2/5\\
                 & $30-40\%$      & --           &$0.095\pm0.004$  & $0.517\pm0.009$  & $4250\pm160$  & $0.0975 \pm0.004$       &$1.2$             & 2/5\\
                 & $40-50\%$      & --           &$0.090\pm0.006$  & $0.512\pm0.010$  & $4000\pm150$  & $0.067 \pm0.004$        &$1.8$             & 5/5\\
                 & $50-60\%$      & --           &$0.086\pm0.005$  & $0.514\pm0.010$  & $3800\pm150$  & $0.049 \pm0.005$        &$2$               & 7/5\\
                 & $60-70\%$      & --           &$0.081\pm0.005$  & $0.513\pm0.011$  & $3610\pm170$  & $0.029 \pm0.004$        &$2$               & 1/5\\
                 & $70-80\%$      & --           &$0.080\pm0.004$  & $0.510\pm0.007$  & $3400\pm176$  & $0.015 \pm0.005$        &$2$               & 4/5\\
\cline{2-8}
 Fig. 1          & $0-5\%$        & $p$          &$0.113\pm0.006$  & $0.490\pm0.010$  & $4700\pm170$  & $0.0165\pm0.003$        &$1.2$             & 33/9\\
   Au-Au         & $5-10\%$       & --           &$0.109\pm0.005$  & $0.500\pm0.011$  & $4530\pm160$  & $0.0094\pm0.0005$       &$1$               & 20/9\\
   62.4 GeV      & $10-20\%$      & --           &$0.105\pm0.004$  & $0.500\pm0.009$  & $4400\pm155$  & $0.0113\pm0.004$        &$1.2$             & 5/9\\
                 & $20-30\%$      & --           &$0.100\pm0.005$  & $0.490\pm0.010$  & $4225\pm140$  & $0.008\pm0.0005$        &$1.3$             & 3/9\\
                 & $30-40\%$      & --           &$0.097\pm0.005$  & $0.488\pm0.011$  & $4160\pm150$  & $0.0055 \pm0.0004$      &$1.5$             & 7/9\\
                 & $40-50\%$      & --           &$0.093\pm0.005$  & $0.489\pm0.009$  & $3900\pm150$  & $0.0035 \pm0.0004$      &$0.8$             & 3/9\\
                 & $50-60\%$      & --           &$0.088\pm0.004$  & $0.487\pm0.011$  & $3700\pm158$  & $0.0022 \pm0.0003$      &$0.6$             & 4/9\\
                 & $60-70\%$      & --           &$0.083\pm0.006$  & $0.480\pm0.012$  & $3530\pm160$  & $0.00175 \pm0.0004$     &$0.3$             & 4/9\\
                 & $70-80\%$      & --           &$0.081\pm0.005$  & $0.481\pm0.009$  & $3310\pm130$  & $0.00055 \pm0.00005$    &$0.4$             & 14/9\\
   \cline{2-8}
   Fig. 1        & $0-5\%$        & $\Lambda$    &$0.137\pm0.006$  & $0.470\pm0.009$  & $4300\pm152$  & $0.023\pm0.004$         &$0.7$             & 1/7\\
   Au-Au         & $5-10\%$       & --           &$0.133\pm0.005$  & $0.468\pm0.010$  & $4120\pm160$  & $0.002\pm0.0004$        &$0.7$             & 1/7\\
   62.4 GeV      & $10-20\%$      & --           &$0.130\pm0.006$  & $0.470\pm0.011$  & $4000\pm187$  & $0.00017\pm0.00004$     &$0.7$             & 1/7\\
                 & $20-30\%$      & --           &$0.126\pm0.004$  & $0.465\pm0.010$  & $3830\pm164$  & $1\times10^{-5}\pm4\times10^{-6}$ &$0.8$   & 1/7\\
                 & $30-40\%$      & --           &$0.123\pm0.004$  & $0.467\pm0.012$  & $3650\pm160$  & $9\times10^{-7}\pm5\times10^{-8}$ &$0.8$   & 1/7\\
                 & $40-60\%$      & --           &$0.120\pm0.005$  & $0.460\pm0.011$  & $3400\pm156$  & $4\times10^{-8}\pm3\times10^{-9}$ &$0.8$   & 5/7\\
                 & $60-80\%$      & --           &$0.117\pm0.005$  & $0.460\pm0.012$  & $3200\pm140$  & $1\times10^{-9}\pm5\times10^{-10}$&$0.9$    & 5/7\\
\cline{2-8}
   Fig. 1        & $0-5\%$        & $\bar \Xi^+$ &$0.138\pm0.004$  & $0.455\pm0.011$  & $4150\pm150$  & $0.0008\pm0.00004$      &$0.7$             & 0.4/5\\
   Au-Au         & $5-10\%$       & --           &$0.134\pm0.005$  & $0.450\pm0.011$  & $4000\pm140$  & $6.5\times10^{-5}\pm6\times10^{-6}$ &$1$   & 3/6\\
   62.4 GeV      & $10-20\%$      & --           &$0.131\pm0.006$  & $0.452\pm0.012$  & $3800\pm157$  & $5.2\times10^{-6}\pm4\times10^{-7}$ &$0.8$ & 2/6\\
                 & $20-40\%$      & --           &$0.127\pm0.004$  & $0.450\pm0.010$  & $3600\pm148$  & $4.5\times10^{-7}\pm6\times10^{-8}$ &$0.7$ & 3/6\\
                 & $40-60\%$      & --           &$0.124\pm0.005$  & $0.453\pm0.010$  & $3400\pm150$  & $8.8\times10^{-9}\pm5\times10^{-10}$ &$0.7$& 1/6\\
                 & $60-80\%$      & --           &$0.120\pm0.005$  & $0.450\pm0.009$  & $3200\pm146$  & $3.4\times10^{-10}\pm5\times10^{-11}$&$0.4$& 3/4\\
\cline{2-8}
 Fig. 1          & $0-20\%$       &$\bar \Omega^+$&$0.138\pm0.004$ & $0.440\pm0.008$ & $4000\pm155$  & $5.2\times10^{-5}\pm5\times10^{-6}$ &$0.6$ & 0.3/0\\
   Au-Au         & $20-40\%$      & --           &$0.134\pm0.006$  & $0.435\pm0.011$  & $3800\pm146$  & $2\times10^{-7}\pm6\times10^{-8}$   &$1$   & 1/0\\
   62.4 GeV      & $40-60\%$      & --           &$0.127\pm0.005$  & $0.436\pm0.012$  & $3600\pm160$  & $3.2\times10^{-9}\pm7\times10^{-10}$ &$0.7$& 2/-1\\
   \cline{2-8}
   Fig. 1        & $0-10\%$       & $d$        &$0.114\pm0.005$  & $0.400\pm0.011$    & $3400\pm140$  & $0.00085\pm0.00005$                 &$1.6$  & 3/7\\
   Au-Au         & $10-20\%$      & --           &$0.109\pm0.006$  & $0.395\pm0.010$  & $3200\pm150$  & $0.00034\pm0.00004$                 &$1.6$  & 2/7\\
   62.4 GeV      & $20-40\%$      & --           &$0.104\pm0.005$  & $0.396\pm0.011$  & $3000\pm145$  & $0.0001\pm0.00004$                  &$1.5$  & 1/7\\
                 & $40-60\%$      & --           &$0.097\pm0.005$  & $0.393\pm0.012$  & $2800\pm170$  & $2\times10^{-5}\pm5\times10^{-6}$   &$1.3$  & 1/7\\
                 & $60-80\%$      & --           &$0.090\pm0.004$  & $0.392\pm0.011$  & $2632\pm150$  & $2\times10^{-6}\pm4\times10^{-7}$   &$0.9$  & 22/6\\
  \hline
   Fig. 2        & $0-5\%$        & $\pi^+$      &$0.130\pm0.005$  & $0.584\pm0.012$  & $7000\pm200$  & $345\pm36    $          & $0.8$            & 89/36\\
   Pb-Pb         & $5-10\%$       & --           &$0.127\pm0.006$  & $0.583\pm0.010$  & $6816\pm191$  & $165.55\pm23$           &$0.7$             & 158/36\\
   2.76 TeV      & $10-20\%$      & --           &$0.123\pm0.004$  & $0.580\pm0.011$  & $6650\pm185$  & $60.55\pm8$             &$0.8$             & 93/36\\
                 & $20-30\%$      & --           &$0.119\pm0.005$  & $0.581\pm0.010$  & $6392\pm180$  & $18.80\pm3$             &$0.9$             & 58/36\\
                 & $30-40\%$      & --           &$0.115\pm0.005$  & $0.580\pm0.012$  & $6200\pm185$  & $6.3 \pm0.4$            &$1$               & 54/36\\
                 & $40-50\%$      & --           &$0.112\pm0.006$  & $0.580\pm0.011$  & $6000\pm170$  & $2.2 \pm0.3$            &$1$               & 92/36\\
                 & $50-60\%$      & --           &$0.109\pm0.004$  & $0.581\pm0.011$  & $5843\pm162$  & $0.66 \pm0.04$          &$1$               & 100/36\\
                 & $60-70\%$      & --           &$0.106\pm0.006$  & $0.579\pm0.010$  & $5670\pm170$  & $0.16 \pm0.03$          &$1.1$             & 197/36\\
                 & $70-80\%$      & --           &$0.101\pm0.005$  & $0.578\pm0.011$  & $5500\pm166$  & $0.04 \pm0.005$         &$1.1$             & 151/36\\
                 & $80-90\%$      & --           &$0.098\pm0.004$  & $0.576\pm0.010$  & $5300\pm160$  & $0.008 \pm0.0004$       &$1.1$             & 221/36\\
\cline{2-8}
 Fig. 2          & $0-5\%$        & $p$          &$0.131\pm0.005$  & $0.570\pm0.010$  & $6700\pm180$  & $8\pm0.7$               &$1$               & 58/30\\
   Pb-Pb         & $5-10\%$       & --           &$0.127\pm0.006$  & $0.570\pm0.010$  & $6500\pm170$  & $4.05\pm0.5$            &$0.9$             & 125/37\\
   2.76 TeV      & $10-20\%$      & --           &$0.123\pm0.005$  & $0.570\pm0.009$  & $6320\pm170$  & $1.35\pm0.3$            &$1.1$             & 37/33\\
                 & $20-30\%$      & --           &$0.120\pm0.006$  & $0.570\pm0.011$  & $6180\pm160$  & $0.9\pm0.05$            &$1.1$             & 34/31\\
                 & $30-40\%$      & --           &$0.116\pm0.005$  & $0.565\pm0.012$  & $6000\pm180$  & $0.16 \pm0.04$          &$1.07$            & 17/31\\
                 & $40-50\%$      & --           &$0.112\pm0.006$  & $0.567\pm0.009$  & $5830\pm170$  & $0.05 \pm0.004$         &$1.1$             & 108/33\\
                 & $50-60\%$      & --           &$0.108\pm0.005$  & $0.564\pm0.010$  & $5650\pm165$  & $0.016 \pm0.003$        &$1$               & 62/31\\
                 & $60-70\%$      & --           &$0.104\pm0.005$  & $0.562\pm0.011$  & $5480\pm170$  & $0.0045 \pm0.0004$      &$1$               & 140/34\\
                 & $70-80\%$      & --           &$0.101\pm0.005$  & $0.562\pm0.009$  & $5300\pm180$  & $0.001\pm0.0003$        &$0.9$             & 214/36\\
                 & $80-90\%$      & --           &$0.097\pm0.004$  & $0.560\pm0.010$  & $5100\pm180$  & $0.0002\pm0.00003$      &$0.8$             & 207/37\\
   \cline{2-8}
   Fig. 2        & $0-10\%$       &$\Lambda$    &$0.155\pm0.005$  & $0.500\pm0.011$  & $6000\pm200$  & $0.13\pm0.03$           &$0.9$             & 28/14\\
   Pb-Pb         & $10-20\%$      & --          &$0.152\pm0.006$  & $0.497\pm0.009$  & $5800\pm180$  & $0.1\pm0.03$            &$0.8$             & 27/14\\
   2.76 TeV      & $20-40\%$      & --          &$0.147\pm0.004$  & $0.496\pm0.010$  & $5600\pm180$  & $0.06\pm0.004$          &$0.8$             & 35/14\\
                 & $40-60\%$      & --          &$0.142\pm0.005$  & $0.495\pm0.011$  & $5400\pm185$  & $0.024\pm0.004$         &$0.6$             & 124/14\\
                 & $60-80\%$      & --          &$0.137\pm0.005$  & $0.494\pm0.010$  & $5200\pm170$  & $0.0074\pm0.0005$       &$1.1$             & 17/14\\
\cline{2-8}
   Fig. 2        & $0-10\%$       & $\Xi$      &$0.156\pm0.006$  & $0.480\pm0.010$  & $5500\pm200$  & $0.0180\pm0.004$        &$1$               & 16/7\\
   Pb-Pb         & $10-20\%$      & --         &$0.152\pm0.005$  & $0.477\pm0.012$  & $5300\pm180$  & $0.0140\pm0.003$        &$1$               & 37/7\\
   2.76 TeV      & $20-40\%$      & --         &$0.148\pm0.005$  & $0.474\pm0.010$  & $5126\pm170$  & $0.0085\pm40.0005$      &$0.9$             & 63/7\\
                 & $40-60\%$      & --         &$0.144\pm0.005$  & $0.473\pm0.009$  & $4950\pm160$  & $0.0032\pm0.0005$       &$0.8$             & 82/7\\
                 & $60-80\%$      & --         &$0.140\pm0.006$  & $0.470\pm0.012$  & $4700\pm180$  & $0.0008\pm0.00005$      &$0.6$             & 107/7\\
\cline{2-8}
 Fig. 2          & $0-10\%$       & $\Omega$   &$0.158\pm0.005$  & $0.460\pm0.010$  & $5000\pm150$  & $0.0014\pm0.0003$       &$1.1$             & 12/2\\
   Pb-Pb         & $10-20\%$      & --         &$0.154\pm0.005$  & $0.458\pm0.009$  & $4817\pm160$  & $9.7\times10^{-4}\pm4\times10^{-5}$&$1.2$  & 1/2\\
   2.76 TeV      & $20-40\%$      & --         &$0.150\pm0.004$  & $0.457\pm0.012$  & $4600\pm180$  & $6\times10^{-4}\pm6\times10^{-5}$  &$0.9$  & 3/2\\
                 & $40-60\%$      & --         &$0.147\pm0.005$  & $0.456\pm0.010$  & $4400\pm120$  & $2\times10^{-4}\pm5\times10^{-5}$  &$0.8$  & 6/2\\
                 & $60-80\%$      & --         &$0.143\pm0.004$  & $0.453\pm0.012$  & $4200\pm180$  & $4\times10^{-5}\pm5\times10^{-6}$  &$0.7$  & 5/1\\
   \cline{2-8}
   Fig. 2        & $0-10\%$       & $d$        &$0.133\pm0.005$  & $0.430\pm0.008$  & $4500\pm170$  & $4.6\times10^{-4}\pm5\times10^{-5}$&$1.8$  & 6/16\\
   Pb-Pb         & $10-20\%$      & --         &$0.130\pm0.005$  & $0.428\pm0.011$  & $4300\pm160$  & $1.8\times10^{-4}4\pm\times10^{-5}$&$1.8$  & 6/16\\
   2.76 TeV      & $20-40\%$      & --         &$0.126\pm0.006$  & $0.426\pm0.009$  & $4100\pm153$  & $6.6\times10^{-5}\pm4\times10^{-6}$&$1.7$  & 4/16\\
                 & $40-60\%$      & --         &$0.122\pm0.004$  & $0.423\pm0.011$  & $3938\pm160$  & $1.5\times10^{-5}\pm5\times10^{-6}$&$1.2$  & 15/10\\
                 & $60-80\%$      & --         &$0.118\pm0.005$  & $0.422\pm0.012$  & $3650\pm150$  & $2.8\times10^{-6}\pm4\times10^{-7}$&$1.3$  & 10/10\\
  \hline
\end{tabular}%
\end{center}}
\end{table*}

Figure 4 is similar to Fig. 3, but shows the dependence of
transverse flow velocity ($\beta_T$) on centrality class and mass
of the particles. One can see that $\beta_T$ depends on the rest
mass of the particles. Greater the mass of the particle is, the
smaller the transverse flow velocity. In fact some hydrodynamic
simulations observed the same velocity for all the particles flow,
but they have different explanations. Besides, different model
give different results. The selection of $\beta_T$ is more
technical and complex, even in some cases it depends on the range
of $p_T$, such that the selection of ranges are different for
different models. Furthermore, there is no centrality dependence
of $\beta_T$ observed in present work, as $\beta_T$ is almost the
same in the central and peripheral collisions.  The reason behind
this is that the collective behavior at the stage of kinetic
freezeout does not change from central to peripheral collisions.
However $\beta_T$ is larger at LHC than that of RHIC.

Figure 5 is similar to Fig. 3 and 4, but shows the dependence of
$V$ on centrality class and mass of the particles. One can see
that $V$ decrease continuously from central to peripheral
collisions because the central collisions correspond to large
number of binary collisions due to the re-scattering of partons
and hence the system with more participants reaches quickly to
equilibrium state. While the number of participants decrease with
the decrease of event centrality and the system goes to
equilibrium state in a steady manner from central to peripheral
collisions. Additionally, $V$ depends on the mass of the
particles. Greater the mass of the particle is, lower the $V$. $V$
at LHC is larger than that at RHIC.

 It should be noted that the situation of $T_0$ and/or $\beta_T$
are very complex on the basis of their dependence on centrality.
The observed results can be changed by changing the model, or the
same model but different method, or by changing the limits and
conditions of the model, such that by changing the parameters, we
can get different results. For example, if for central collisions,
one use a smaller $T_0$ and a larger $\beta_T$, a decreasing trend
for $T_0$ from peripheral to central collisions can be obtained.
At the same time, a negative correlation between $T_0$ and
$\beta_T$ will also be obtained. Similarly if one use a larger
$T_0$ and a smaller $\beta_T$, an increasing trend for $T_0$ from
peripheral to central collisions can be obtained. At the same
time, a positive correlation between $T_0$ and $\beta_T$ will also
be obtained.
\\
 {\section{Conclusions}}
 The main observations and conclusions are summarized here.

 a) The transverse momentum spectra of different particle species are analyzed by the blast wave model with Boltzmann Gibbs statistics and the bulk properties in terms of the kinetic freezeout temperature, transverse flow velocity and freezeout volume are extracted in different centrality classes in nucleus-nucleus collisions at center of mass energy.

 b) It is observed that $T_0$ is dependent on the cross-section of the interacting particle, i:e Larger the production cross-section of the interacting particle correspond to smaller $T_0$.

 c) A double kinetic freezeout scenario is observed due to the separate decoupling of non-strange and strange (multi-strange) particles.

 d) The transverse flow velocity ($\beta_T$)  and kinetic freezeout volume ($V$) are observed to depend on the mass of the particles, i:e. The larger the mass of the particle, the smaller $\beta_T$ and $V$.

 e) Kinetic freezeout temperature ($T_0$) and freezeout volume ($V$) decrease from central peripheral collisions due to decrease of degree of excitation of interacting system and the decrease of the number of binary collisions due to the re-scattering of partons respectively from central to peripheral collisions. While $\beta_T$ is observed to be independent of centrality and remains almost unchanged from central to peripheral collisions because the collective behavior does at the stage of kinetic freezeout in interacting system does not change with event centrality.

f) $T_0$, $\beta_T$ and $V$ are observed to be larger at LHC
collisions that at RHIC.

g)  The obtained results can be changed by changing the model, or
the same model with different method or by changing the parameters
used in the model.

   )
\\
\\

{\bf Data availability}

The data used to support the findings of this study are included
within the article and are cited at relevant places within the
text as references.
\\
\\
{\bf Compliance with Ethical Standards}

The authors declare that they are in compliance with ethical
standards regarding the content of this paper.
\\
\\
{\bf Acknowledgements}

The authors would like to thank support from the National Natural Science
Foundation of China (Grant Nos. 11875052, 11575190, and 11135011).
\\
\\

\end{multicols}
\end{document}